\documentclass[aps,pra,preprint,superscriptaddress,longbibliography]{revtex4-1}

\usepackage{amsmath}
\usepackage{amsthm}
\usepackage[english]{babel}
\usepackage{amsfonts}
\usepackage{amssymb}
\usepackage{xcolor,graphicx}
\usepackage{tikz}
\usepackage{color}
\usepackage[pdftex]{hyperref} 
\usepackage{longtable}
\usepackage{array}
\usepackage{dsfont}

\begin{document}
	
	\title{Spectral Collapse in the two-photon quantum Rabi model}

	\author{R. J. Armenta Rico}
	\affiliation{Facultad de Ciencias F\'isico - Matem\'aticas, Universidad Aut\'onoma de Sinaloa, Calle Universitarios Ote., Cd Universitaria, Universitaria, 80010 Culiac\'an Rosales, Sin.}
	
	\author{F. H. Maldonado-Villamizar}
	\email[e-mail: ]{fmaldonado@inaoep.mx}
	\affiliation{CONACYT-Instituto Nacional de Astrof\'isica, \'Optica y Electr\'onica, Calle Luis Enrique Erro No. 1, Sta. Ma. Tonantzintla, Pue. CP 72840, M\'exico}

	\author{B. M. Rodriguez-Lara}
	\affiliation{Tecnologico de Monterrey, Escuela de Ingeniería y Ciencias, Ave. Eugenio Garza Sada 2501, Monterrey, N.L., 64849, M\'exico}
	\affiliation{Instituto Nacional de Astrof\'isica, \'Optica y Electr\'onica, Calle Luis Enrique Erro No. 1, Sta. Ma. Tonantzintla, Pue. CP 72840, M\'exico}	
	
	\begin{abstract}
	Spectral collapse, the transition from discrete to continuous spectrum, is a characteristic in quantum Rabi models.
	We explore this phenomenon in the two-photon quantum Rabi model in optical phase space and find that, in the so-called degenerate qubit regime, the collapse is similar to that happening in the transition from a quantum harmonic to an inverted quadratic potential with the free-partical potential as transition point.
	In this regime, it is possible to construct Dirac-normalizable eigenfunctions for the model that show well defined parity.
	In the general model, we use parity to diagonalize the system in the qubit basis and numerically find that the qubit frequency does not change the critical point where spectral collapse occurs. 
	\end{abstract}
	
	
	\maketitle
\section{Introduction}

The quantum Rabi model \cite{Braak2011p100401} describes the minimal coupling between a two-level system and a boson field.
Experimental realizations of the model exist in a range of quantum platforms \cite{Wallraff2004p162,Niemczyk2010p772,Yoshihara2016p44}.
Experimentalists can tailor their quantum systems to explore extensions of the model \cite{Clarke2008p1031} that has fueled theoretical extensions as well \cite{Moroz2016p50004,Wang2014p54001,Guan2018p315204,Rodriguez2018p043805,Chilingaryan2013p335301,Chilingaryan2015p245501}.
We are interested in the process of two-photon exchange reported with atoms \cite{Bertet_2002p143601} and solid state devices \cite{Stufler_2006p73125304}.
In this so-called two-photon quantum Rabi model,
\begin{equation}\label{eq:001}
H=\frac{\omega_0}{2}\hat{\sigma}_{z} + \omega a^{\dagger}\hat{a} + g_2(\hat{a}^{\dagger 2}+\hat{a}^{2})\hat{\sigma}_{x},
\end{equation}
the qubit and boson field are described by the two-level energy gap related to the frequency $\omega_0$ and the frequency $\omega$.
The coupling between these is given by the parameter $g_{2}$.
The qubit is described by Pauli matrices $\hat{\sigma}_{i}$ and the field by annihilation (creation) operators $\hat{a}$ ($\hat{a}^{\dagger}$).
In the two-photon quantum Rabi model, qubit flop accompanies the creation or destruction of two photons.
This process resembles parametric up (down) conversion .
The latter has varied theoretical and experimental applications \cite{Bertet_2005p257002,Felicetti_2015p033817,Puebla_2017p063844,Cui_2017p204001,Felicetti_2018p013851,Pedernales_2018p160403,Lupo_2019p4156}, that might point to the importance of a better understanding of the former.

An interesting characteristic of the single-photon quantum Rabi model is the spectral collapse that occurs in the so-called relativistic regime \cite{Maldonado2019p013811}. 
There, we can follow the transition from a discrete spectrum in the so-called degenerate qubit regime where the model is equivalent to a driven cavity and the so-called relativistic regime where the model is equivalent to Dirac equation in $(1+1)$D with continuous spectrum.
The eigenstates of the model interpolating between these two regimes transition from the superposition of even and odd displaced number states to that of infinitely squeezed coherent states, respectively. 
Here, we want to show that the  spectral collapse in the two-photon quantum Rabi model, existing at a critical coupling and providing a spectrum with discrete and continuous part \cite{Duan2016p464002}, has a different nature than the one in the single-photon quantum Rabi model.

In the following, we discuss the spectral collapse mechanism in the two-photon quantum Rabi model.
First, we rewrite the Hamiltonian for the model in optical phase space. 
This allows us to compare it with a quantum mechanical system whose effective potential transitions from harmonic to inverted oscillator form.
Second, we analytically study this transition in the degenerate qubit regime and provide its solution in quadrature-representation that confirms the critical coupling dependence on just the boson frequency. 
Then, we conduct a numerical analysis in both the full two-photon quantum Rabi Hamiltonian and in its diagonalization in the qubit basis to confirm the null effect of the qubit frequency on the critical coupling.
We present results for on-resonance and off-resonance surveys that also confirm the existence of an exceptional eigenstate with finite norm at the critical coupling of every parameter set.
Finally, we close with a conclusion.

\section{Optical phase space model}

We are interested in the mechanism behind spectral collapse reported in the two-photon quantum Rabi model \cite{Duan2016p464002,Duan2019p18352}. 
For reasons that will become clear in the following, we move into optical phase space \cite{Mandel_wolf_1995}, 
\begin{eqnarray}
	\hat{q}= \frac{1}{\sqrt{2}} \left( \hat{a}^{\dagger} + \hat{a}  \right), \qquad \hat{p}= \frac{ i }{\sqrt{2}} \left( \hat{a}^{\dagger} - \hat{a} \right),
\end{eqnarray} 
and perform a $\pi/2$ rotation around the axis defined by $\hat{\sigma}_y$ such that the we arrive to the Hamiltonian, up to a constant factor,
\begin{eqnarray}
\hat{H}_{y} = \frac{1}{2} \left[ \omega +  2 g_{2} \hat{\sigma}_{z} \right] \hat{p}^{2} +  \frac{1}{2} \left[ \omega - 2 g_{2} \hat{\sigma}_{z}  \right] \hat{q}^{2}  + \frac{1}{2} \omega_{0} \hat{\sigma}_{x} .
\end{eqnarray}
Moving into a frame defined by a unitary rotation in terms of a Fourier-like rotation,
\begin{eqnarray}
\hat{U}(\theta) = \left[ \hat{R}(\theta) \right]^{-\frac{1}{2} \left(\hat{\sigma}_{z} - 1 \right)}, \qquad \hat{R}(\theta) = e^{ -\frac{i\theta}{2} \left(\hat{p}^{2} + \hat{q}^{2} \right)},
\end{eqnarray}
and choosing a rotation angle $\theta = \pi/2$, we recover a rotated two-photon quantum Rabi model Hamiltonian,  
\begin{equation}\label{eq:001b}
\hat{H}_{R} = \frac{1}{2} \left(
\alpha_{+} \hat{p}^{2} +  \alpha_{-} \hat{q}^{2} \right) \hat{\sigma}_{0} + \frac{1}{2} \omega_{0} \left[\hat{R}_{\frac{\pi}{4}}  \hat{\sigma}_{+} + \hat{R}^{\dagger}_{\frac{\pi}{4}}  \hat{\sigma}_{-} \right],
\end{equation}
where we define the dimensionless auxiliary parameters $\alpha_{\pm}= \omega \pm 2g_{2} $ and use the shorthand notation $\hat{\sigma}_{0}$ for the identity matrix and $\hat{R}_{\theta} \equiv \hat{R}(\theta)$.
This analogy immediately brings to our mind the idea of spectral collapse as the diagonal element of this Hamiltonian transitions from harmonic to free-particle form when the auxiliary parameter takes the value $\alpha_{-} = 0$ at the critical coupling $g_{c} = \omega / 2$.
For values larger than the critical coupling, the diagonal element takes the form of an inverted oscillator, $\alpha_{-} < 0$.

In the past, we showed that a competition between Hamiltonians with components showing discrete and continuous spectrum produces the spectral collapse in the single-photon quantum Rabi model \cite{Maldonado2019p013811}.
There, the collapse occurs from the transition from driven-cavity-like, in the degenerate-qubit regime, into a relativistic $(1+1)$D Dirac-like Hamiltonian, in the relativistic regime.
We have a different mechanism in the two-photon quantum Rabi model.
Here, the diagonal term shows spectral collapse in the degenerate-qubit regime equivalent to a transition from harmonic oscillator to free-particle and then to inverted harmonic oscillator.

\section{Degenerate qubit regime}

\begin{widetext}
\begin{figure}[htb!]
	\includegraphics[scale=1]{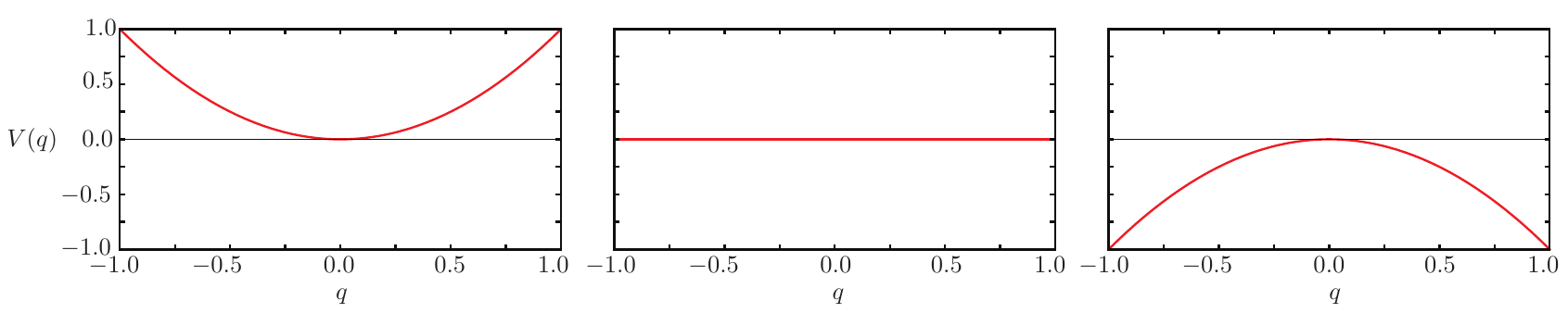}
	\caption{Effective pseudo-potential $V(\hat{q}) = (\omega - 2 g_{2}) \hat{q}^{2}$ in the diagonal terms of the rotated two-photon quantum Rabi model with (a) $\omega > 2 g_{2}$, (b) $\omega = 2 g_{2}$ and (c) $\omega < 2 g_{2}$.}
	\label{fig:Fig1}
\end{figure}
\end{widetext}

Let us focus on the boson component attached to the identity element in the qubit basis, an analogy to the so-called degenerate qubit regime where $\omega_{0} \rightarrow 0$,
\begin{eqnarray}
\hat{H}_{0} = \frac{1}{2} \left(
\alpha_{+} \hat{p}^{2} + \alpha_{-} \hat{q}^{2} \right).
\end{eqnarray} 
We can think of the second right-hand-side term as a potential $V(\hat{q}) = \alpha_{-} \hat{q}^{2}$ that takes the form of a harmonic oscillator showing discrete spectrum for $\alpha_{-} >0$ with $\omega > 2 g_{2}$, Fig. \ref{fig:Fig1}(a), and two regimes with continuous spectrum in the form of free-particle for $\alpha_{-} = 0$ with $\omega = 2 g_{2}$, Fig. \ref{fig:Fig1}(b), or inverted oscillator for $\alpha_{-} <0$ with $\omega < 2 g_{2}$, Fig. \ref{fig:Fig1}(c). 
Thus, the spectral collapse in the degenerate-qubit regime is related to the transition from harmonic to free-particle potential.

Suppose that the vector $\vert \lambda_{0} \rangle$ is an eigenvector of the diagonal element $\hat{H}_{0}$ with eigenvalue $\lambda_{0}$.
We can use a quadrature-representation to find linearly independent solutions to the diagonal element,
\begin{widetext}
\begin{align}
\langle q\vert \lambda_{0}\rangle &= c_1 \, e^{-\frac{1}{4}\alpha q^2} \,_1F_1\left(-\nu-\frac{1}{4};\frac{1}{2};\frac{1}{2} \alpha q^2 \right) + c_2\, q  \, e^{-\frac{1}{4}\alpha q^2} \, _1F_1\left(-\nu+\frac{1}{4};\frac{3}{2};\frac{1}{2} \alpha q^2
\right),
\end{align}
\end{widetext}
in terms of the confluent hypergeometric function $_1F_{1}\left(a;b;z\right)$ \cite{Lebedev1972p}.
We introduce the auxiliary parameters $\alpha^{2} = (\omega - 2 g_{2}) / (\omega + 2 g_{2})$ and $\Omega=\sqrt{\omega^{2}-4g_{2}^{2}}$, modal amplitudes $c_{1}$ and $c_{2}$ that include normalization constants, and the scaled eigenvalue 
\begin{eqnarray}
\nu =   \frac{\omega}{\Omega} \lambda - \frac{1}{2},
\end{eqnarray}
that is real for parameters $\omega > 2g_{2}$, diverges for $\omega = 2g_{2}$, and becomes complex for $\omega < 2g_{2}$.
In the first case $\omega > 2g_{2}$, we recover the discrete, equidistant, harmonic oscillator spectrum,
\begin{eqnarray}
\lambda_{n}^{(I)}=\frac{\Omega}{\omega}\left(n+\frac{1}{2}\right), \qquad n=0,1,2,\ldots
\end{eqnarray}
with corresponding Hermite-Gauss eigenfunctions,
\begin{align}
\langle q\vert \lambda_{n}^{(I)} \rangle &=\frac{1}{\sqrt{2^{n} n!}} \left(\frac{\alpha}{\pi}\right)^{\frac{1}{4}} e^{-\frac{1}{2} \alpha q^2} H_{n} (\alpha q).
\end{align}
Then,  for the free-particle-like case $\omega = 2g_{2}$, the spectrum collapse and becomes continuous with Dirac-delta normalizable forward and backward propagating monocromatic plane wave eigenfunctions,
\begin{equation}\label{eq:007g}
\langle q\vert  \pm \lambda^{(II)} \rangle= \frac{1}{\sqrt{2 \pi}} e^{ \pm i\sqrt{\lambda^{(II)}} q}, 
\end{equation}
with real positive eigenvalues $\lambda^{(II)} \in \left[0, \infty \right)$.
In the inverted oscillator case $\omega < 2 g_{2}$, the spectrum remains continuous and the eigenfunctions are Dirac-delta normalizable \cite{WolfKB2010p83},
\begin{align}\label{eq:full}
\langle q\vert \lambda^{III}\rangle &=\frac{e^{-i\frac{\pi}{4}\left(\eta+1\right)}\Gamma\left(-\eta\right)}{2^{\frac{3}{4}}\pi} D_{\eta}\left(\pm \left\vert\frac{ \omega-2g_{2}}{\omega+2g_{2}}\right\vert^{\frac{1}{4}} e^{3i\frac{\pi}{4}}\sqrt{2}q\right),
	\end{align}
\noindent
with real eigenvalue $\lambda^{(III)} \in \mathbb{R}$, where we define the auxiliary real parameter $\eta= i \lambda^{(III)} \Omega^{-1} - 1/2 $, and  use the parabolic cylinder function $D_{\eta}(x)$ \cite{Lebedev1972p}.
We want to stress that both Dirac-normalizable solutions with continuous spectra allow the construction of states with well defined parity.

\section{General model}

Now, let us try to address the process behind the spectral collapse in the general model by considering its symmetries \cite{Duan2016p464002}.
For starters, we can partition the boson Hilbert space into even and odd sectors and rewrite the rotated two-photon quantum Rabi Hamiltonian, up to a constant,
\begin{eqnarray}
H_{q} = \frac{\omega_0}{2}\hat{\sigma}_{x} +2 \omega \hat{K}_{z} - 2 g_2 \left( \hat{K}_{+} + \hat{K}_{-} \right) \hat{\sigma}_{z},
\end{eqnarray}
in terms of the elements of the $SU(1,1)$ group such that $\left[ \hat{K}_{z}, \hat{K}_{\pm} \right] = \pm \hat{K}_{\pm}$ and $\left[ \hat{K}_{+}, \hat{K}_{-} \right] = - 2 \hat{K}_{z}$.
The parameter $q$ is known as Bargmann index and takes the value of $q=1/4$ ($q=3/4$) in the even (odd) boson subspace defined as $ \mathcal{H}_{1/4} = \left\{ \vert 1/4 ; m \rangle\right\}$ ($ \mathcal{H}_{3/4} = \left\{ \vert 3/4 ; m \rangle \right\}$).
Each subspace has a parity operator $\hat{\Pi}_{q} = e^{i \pi(\hat{K}_{z} - q)}$ and the action of these operators in the subspaces can be found in \cite{Linblad1970p27,Groenevelt2001p65}.
We use a Foulton-Gouterman transformation \cite{Moroz2016p50004} to diagonalize the rotated two-photon quantum Rabi Hamiltonian in the qubit basis, 
\begin{eqnarray}
\hat{H}_{FG} = \hat{H}_{q,+} \vert + \rangle \langle + \vert + \hat{H}_{q,-} \vert - \rangle \langle - \vert .
\end{eqnarray}
We use the parity $\hat{\Pi}_{q}$ as auxiliary operator for this.
The four Hamiltonians, one for each boson sector and up to a constant, 
\begin{eqnarray}
\hat{H}_{q, \pm} =   \pm \frac{\omega_0}{2} \hat{\Pi}_{q} +  2\omega \hat{K}_{z} - 2 g_2 \left( \hat{K}_{+} + \hat{K}_{-} \right),
\end{eqnarray}
have a form where the first two right-hand-side terms have discrete and the last term has continuous spectrum.
The competition between these terms defines the spectral collapse in the full rotated two-photon quantum Rabi model.

\begin{figure}[htb!]
	\includegraphics[scale=1]{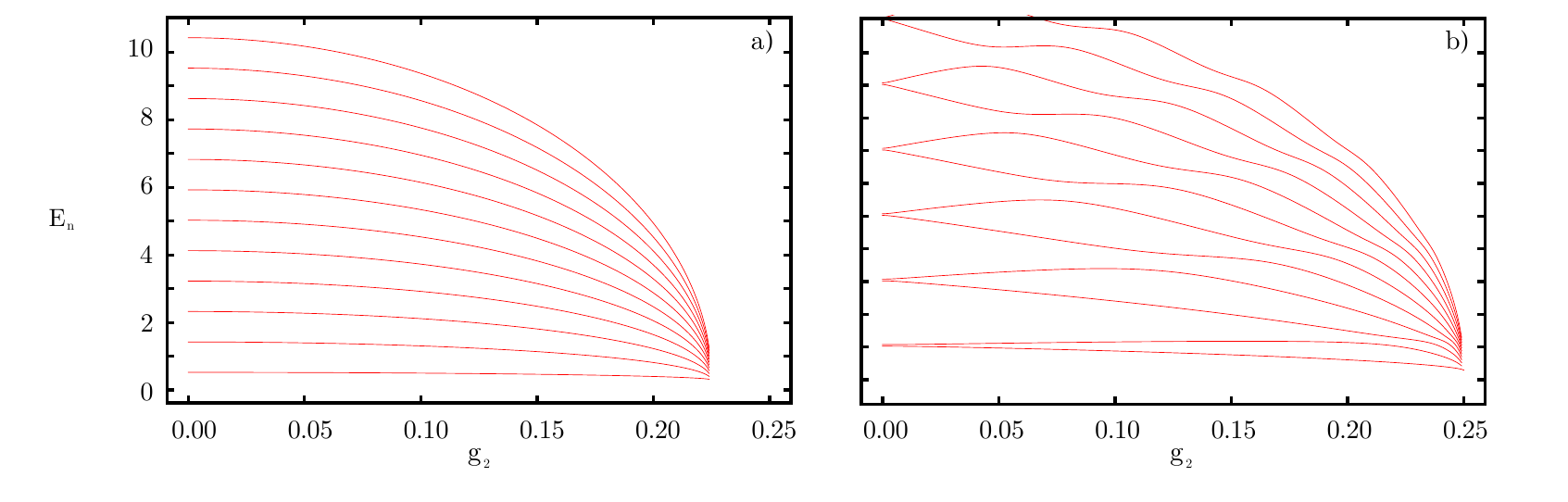}
	\caption{Spectral collapse in the boson sector $\hat{H}_{1/4,+}$ of the rotated two-photon quantum Rabi model diagonalized in the qubit basis in (a) the degenerate qubit regime $\omega_{0}=0$ , $\omega=0.45$ and $g_{2c}=0.225$; (b) on-resonance $\omega_{0} = 1$, $ \omega=0.5$ and $g_{2c}=0.25$.}
	\label{fig:Fig2}
\end{figure}

We can numerically explore this transition in each boson subspace, but we will focus on the even excited subspace $\mathcal{H}_{1/4,+}$ for the sake of brevity. 
Figure \ref{fig:Fig2} shows the first twenty five eigenvalues for this subspace associated to the upper diagonal term $\hat{H}_{1/4,+}$ using a truncated boson Hilbert subspace of $2^{13}$.
We can compare these results with those from the full rotated two-photon quantum Rabi model using a truncated space of $2^{12}$ photons to good agreement. 
In both cases, we accept an eigenvalue and eigenfunction pair if the norm of the last $20\%$ components of the boson sector of the eigenvector is less than $10^{-6}$.
Once we reach the critical value for the coupling constant $g_{c} = 2 \omega$, there is but a single converged eigenfunction as the spectrum becomes continuous and the truncation method is no longer viable to solve the eigenvalue problem.
Analytic and numeric results are in good agreement.
A key characteristic arises in these results, there is always an exceptional solution at the critical coupling $g_{c} =  \omega/{2}$ in each boson subspace as reported in Ref. \cite{Duan2016p464002}.

\begin{figure}[!htb]
	\includegraphics[scale=1]{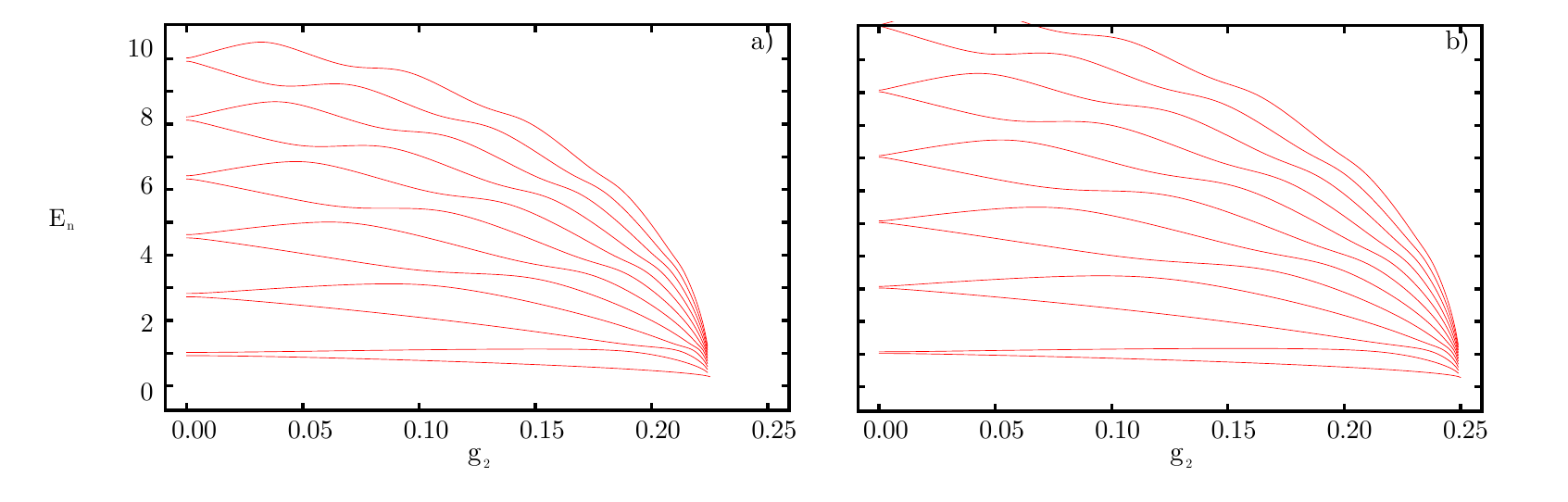}
	\caption{ Spectral collapse in the boson sector $\hat{H}_{1/4,+}$ of the rotated two-photon quantum Rabi model diagonalized in the qubit basis in the off-resonance case with (a) $\omega_{0} = 1$, $\omega = 0.45$ and $g_{2c}=0.225$; (b) $\omega_{0} = 0.95$, $\omega = 0.5$ and $g_{2c}=0.25$.}
	\label{fig:Fig3}
\end{figure}

We explore different parameter space to numerically verify the dependence of the critical coupling on just the field frequency. 
In particular, we surveyed two off-resonance models. 
One with fixed qubit frequency $\omega_{0} = 1$ and variable boson frequency $\omega \in \left[0.45, 0.55\right] \omega_{0}$.
Figure \ref{fig:Fig3}(a) shows a result of this survey for $\omega = 0.45$ that yields a critical coupling of $g_{c} = 0.225$.
Another with qubit frequency $\omega_{0} \in \left[0.95,1.05\right] \omega$ with boson frequency $\omega = 0.5$ that yields $g_{c} = 0.25$. 
Figure \ref{fig:Fig3}(b) shows a result of this survey for $\omega_{0} = 0.95$.
The surveys explored homogeneous twenty steps distributions in the variable frequencies and two hundred steps in the coupling parameter. 
Finer combs in the coupling parameter were implemented centered on the critical coupling $ g_{2} \in \left[0.98,1.02\right] g_{c}$ and covering two hundred steps to verify the results.
It seems that the addition of the parity has no effect on the critical coupling nor on the exceptional solution at the critical coupling.

\section{Conclusions}

We propose that the mechanism behind spectral collapse, the transition from discrete to continuous spectrum, in the two-photon quantum Rabi model is analogous to a transition from a harmonic into an inverted oscillator with the critical point being an analogous to a free particle at a critical coupling parameter of half the boson field frequency. 
It is straightforward to show this mechanism in the degenerate qubit regime using optical phase space representation.
This spectral collapse mechanism remains unchanged outside the degenerate qubit regime as numerical experiments show no change in the critical coupling for the explored values of the qubit frequency.
Our results confirm the existence of an exceptional solution at the critical coupling that aligns with the ground state of the boson subspaces in the regions with discrete spectra.

\begin{acknowledgments}
J.R.A.M acknowledges funding from AMC Programa Verano de la Investigación Científica 2018, F.H.M.-V. from CONACYT C\'{a}tedra Grupal \#551, and B.M.R.-L. from CONACYT grant CB-2015-01 \#255230 and Marcos Moshinsky Foundation Young Researcher Chair 2018.
\end{acknowledgments}


\providecommand{\noopsort}[1]{}\providecommand{\singleletter}[1]{#1}%
\end{document}